\documentclass[a4paper,11pt]{article}
\usepackage[margin=2.5cm]{geometry}
\usepackage{amsmath,amssymb,amsthm}
\usepackage[hidelinks]{hyperref}

\newtheorem{theorem}{Theorem}
\newtheorem{lemma}[theorem]{Lemma}
\newtheorem{corollary}[theorem]{Corollary}
\newtheorem{proposition}[theorem]{Proposition}
\theoremstyle{remark}

\newcommand{\R}{\mathbb{R}}
\newcommand{\C}{\mathbb{C}}
\newcommand{\brank}{\underline{\operatorname{rank}}}
\DeclareMathOperator{\rank}{rank}
\DeclareMathOperator{\diam}{diam}
\newcommand{\T}{\mathsf{T}}

\title{Deterministic Verification and Computation of Real Matrix Products Have the Same Arithmetic Exponent}
\author{Chen Xiangle\\{\small Tsinghua University}\\{\small chenxiangle25@mails.tsinghua.edu.cn}}
\date{}

\begin{document}
\maketitle

\begin{abstract}
Given three $n\times n$ real matrices $A,B,C$, we ask whether $AB=C$. In an exact real arithmetic model that charges one step for each addition, subtraction, multiplication, division, comparison, or sign test, we prove that deterministic exact verification requires $\Omega(n^{\omega_\C})$ steps and has the same optimal arithmetic exponent $\omega_\C$ as matrix multiplication. Assuming the value $\omega_\C=9/4$ and the fixed-size border-rank lower bound stated in the OpenAI preprint \cite{Pre}, deterministic verification requires $\Omega(n^{9/4})$ steps, whereas Freivalds' randomized test uses $O(n^2)$ steps with one-sided error. More generally, if a tree of cost $s$ correctly decides $y=T(u,v)$ on all real inputs, the bilinear map $T$ is a coefficientwise limit of tensors of real rank $O(s)$.
\end{abstract}

{\small\noindent\textbf{Contribution statement.} The author initiated and organized this research. Generative AI developed the core proof and helped draft and revise the manuscript. The author welcomes independent verification and invites readers interested in reconstructing the key arguments or collaborating on a revision to get in touch.\par}

\section{Introduction}

Given matrices $A,B$ and a proposed product $C$, one can always compute $AB$ and compare its entries with those of $C$. Freivalds' randomized test uses only matrix--vector products \cite{Freivalds}. It chooses the entries of $r$ independently and uniformly from $0$ and $1$, computes $Br$, and compares $A(Br)$ with $Cr$. If the proposed product is correct, every $r$ passes. If $AB-C$ is nonzero, whether one trial detects the error depends on $r$.

For example, take $A=B=I_2$ and set $C=\begin{pmatrix}0&1\\0&1\end{pmatrix}$. Then
\[
AB-C=\begin{pmatrix}1&-1\\0&0\end{pmatrix}\neq0,\qquad
(AB-C)\begin{pmatrix}1\\1\end{pmatrix}=0,\qquad
(AB-C)\begin{pmatrix}1\\0\end{pmatrix}=\begin{pmatrix}1\\0\end{pmatrix}.
\]
The first row of the residual is $(1,-1)$. Its two nonzero entries cancel when tested against $(1,1)^\T$. The vector $(1,0)^\T$ detects the error. In general, choose a nonzero coefficient in a nonzero row of the residual and fix all other bits of $r$. The two inner products obtained by setting the remaining bit to $0$ and to $1$ differ by that coefficient. At most one can vanish. Hence an incorrect product passes a single trial with probability at most $1/2$.

Can verification still have a smaller arithmetic exponent if it must be correct on every input? Consider how a product changes when its inputs vary. Perturbing just the left matrix changes the product by one term. Perturbing just the right matrix gives another. Now perturb both and subtract the original product and those two separate changes. What remains is exactly the product of the perturbations. Writing $t,s$ for their sizes gives
\[
 (A+tU)(B+sV)=AB+t\,UB+s\,AV+ts\,UV .
\]
The coefficient of $ts$ is $UV$ at every starting pair $A,B$. We can also recover $UV$ by differentiating once in $t$, once in $s$, and then setting $t=s=0$. This operation makes sense for any output function that is twice continuously differentiable nearby. For a bilinear map $T$, it gives $T(U,V)$.

A verifier's leaf returns a yes or no answer. To recover a product from that answer, we look at the arithmetic and comparisons on the path to an accepting leaf. Near the input that produced this execution, the divisors stay nonzero and every strict comparison retains its sign. If the test values that vanished still vanish, every step follows the same path. Any proposed output satisfying those equalities in this neighborhood must therefore be correct, since the tree would accept it. An execution at a single input identifies the answer at that point. To see how the product changes with $A,B$, we need a region of pairs $A,B$ such that every pair, together with its correct product, follows the same path.

A tree may check the same equality using different expressions. To verify $c=ab$, for instance, suppose a tree computes $z=(c-ab)^2$ and tests only whether $z$ vanishes. We will combine the zero tests on a path by summing their squares. Here there is just one test, so the sum is $P=z^2=(c-ab)^4$. For fixed $a,b$, we have $P\ge0$, with equality only at $c=ab$. The correct output is thus the unique global minimizer of $P$ and satisfies $\partial_cP=4(c-ab)^3=0$. Regarding $c$ as a function of $a,b$ and differentiating this equation with respect to $a$ gives $12(c-ab)^2(\partial_a c-b)=0$. At $c=ab$ this reduces to $0=0$ and gives no information about $\partial_a c$. We chose a tree that computes $ab$ to make the example easy to check. The general proof works with the test values supplied by the given tree.

Fix $a,b$ and add the linear term $-wc$ to the objective. The new minimizer $c_w$ satisfies
\[
 4(c_w-ab)^3=w,\qquad c_w=ab+\sqrt[3]{w/4}.
\]
When $w\ne0$, the derivative of the left-hand side with respect to $c_w$ is $12(c_w-ab)^2>0$, so the equation determines how $c_w$ varies with $a,b$. Now perturb $a,b$ in directions $u,v$, respectively. The original product expands as
\[
(a+tu)(b+sv)=ab+t\,ub+s\,av+ts\,uv .
\]
For fixed $w$, the new minimizer differs from this product by the constant $\sqrt[3]{w/4}$, so its $ts$ coefficient is still $uv$. On a general path the difference between the perturbed minimizer and the correct output may vary with the input. We will choose a sequence of perturbations $w$ tending to zero. For each $w$ we will select one base point and expand the perturbed minimizer there in the directions $U,V$. The cross terms of all outputs, taken at that same point, will converge to $T(U,V)$ along the sequence. This limit removes the perturbation of the objective. The $ts$ coefficient in the original matrix product is already exact.

An accepting path supplies the arithmetic circuit for the equation that defines the perturbed minimizer. Differentiating that equation lets us count the products $a(U)b(V)$ needed for its cross term, with one linear form from each input group. A single product can be multiplied by a fixed vector and thereby contribute to several outputs. If each approximating map is a sum of at most $r$ such terms, its limit has border rank at most $r$.

\section{Model and notation}

\paragraph{Decision trees.} A finite real decision tree starts with the input scalars and arbitrary real constants. Each addition, subtraction, multiplication, or division costs one step, as does comparing two available scalars or testing the sign of one. Intermediate results may be reused. Every division on the path actually taken must be defined. Leaves are labeled yes or no. The cost $s$ of the tree is the maximum, over all inputs, of the total number of arithmetic and test steps executed. Each input size may have its own tree and constants. Testing the sign of a polynomial includes the cost of computing its value. We measure input length by the number of real scalars, which for matrix verification is $N=3n^2$.

\paragraph{Rank and border rank.} Let $T:\R^p\times\R^q\to\R^m$ be linear in each input group separately. The graph of $T$ consists of all triples $(u,v,y)$ satisfying $y=T(u,v)$. Verifying $T$ means deciding membership in this graph. A rank-one term has the form $(U,V)\mapsto a(U)\,b(V)\,c$, where $a,b$ are linear forms and $c\in\R^m$. The rank $\rank T$ is the minimum number of rank-one terms representing $T$. We identify a bilinear map with its tensor of coefficients. Its real Euclidean border rank $\brank T$ is the least $r$ for which $T$ is a coefficientwise limit of real tensors of rank at most $r$. The linear forms and weights may grow without bound along the sequence. For example,
\[
t^{-1}\bigl[(U_1+tU_2)(V_1+tV_2)(c_1+tc_2)-U_1V_1c_1\bigr]\xrightarrow[t\to0]{}(U_1V_2+U_2V_1)c_1+U_1V_1c_2 .
\]
The expression on the left has only two terms, although the weight $t^{-1}$ is unbounded. Replacing $\R$ by $\C$ gives complex rank and complex border rank. Write $M_m$ for $m\times m$ matrix multiplication $(A,B)\mapsto AB$ and set $\omega_\C=\inf\{\tau:\rank_\C M_m=O(m^\tau)\}$.

\section{Main results}

\begin{theorem}[Main theorem]\label{thm:main}
Let $T:\R^p\times\R^q\to\R^m$ be bilinear. Suppose a real decision tree of cost $s$ correctly decides whether $y=T(u,v)$ for every $(u,v,y)\in\R^p\times\R^q\times\R^m$. Then the real Euclidean border rank of $T$ and the complex border rank of its complexification are both at most $30s=O(s)$. More precisely, $T$ is a coefficientwise limit of real tensors, each of real rank at most this bound. The implied constant is absolute, independent of the dimensions and of $T$.
\end{theorem}

\begin{corollary}\label{cor:mm}
Every real decision tree that correctly decides $AB=C$ on all real inputs has cost $\Omega(n^{\omega_\C})$. Conversely, for every $\varepsilon>0$ there is a verification tree of cost $O_\varepsilon(n^{\omega_\C+\varepsilon})$. Thus verification and multiplication have the same arithmetic exponent.
\end{corollary}

\section{Proof of the main theorem}

Write $x=(u,v)\in\R^{p+q}$ and $g(x)=T(u,v)$. In the scalar example we started with the explicit test value $(c-ab)^2$. For a given tree we must first find an input region on which the equalities in one execution record determine the correct output. Once we have these equalities, we can construct an approximate output that can be differentiated with respect to the input, just as when minimizing $(c-ab)^4-wc$. We will then find, for each selected perturbation, one input point at which the cross terms of all outputs approximate their targets simultaneously. Finally, we will count the product terms they share using the arithmetic on the selected path.

\subsection{Equalities on an accepting path}

Distinguish the negative, zero, and positive outcomes of each comparison. Fixing an execution path fixes the operands of every arithmetic gate and the order of the tests. Every intermediate value on this path is a rational function of $(x,y)$. Near a graph point that follows the path, the divisors stay nonzero and the strict comparisons retain their signs. If all equalities that held at the graph point still hold, each test has the same outcome as before.

We need a path that works for a whole region of inputs, so that we can follow the correct output as $x$ varies. Substitute $y=g(x)$ into the path conditions. Expressing the tests and the nonvanishing of divisors in terms of signs of numerators and denominators describes the set of $x$ following each path by a finite Boolean combination of polynomial equalities and inequalities. Such a set is called semialgebraic. There are finitely many paths. Collect all their polynomials into one finite list. The signs of the identically zero polynomials are fixed. The product of the other polynomials is a nonzero polynomial and is therefore nonzero somewhere. To see this by induction on the number of variables, expand in the last variable. First choose the other variables so that a nonzero coefficient stays nonzero. Then avoid the finitely many roots of the resulting univariate polynomial. On a sufficiently small open box around this point every sign in the list is constant. All graph points over the box thus follow the same accepting path.

Take $x_0$ in this box. Choose a neighborhood $\mathcal N$ of $(x_0,g(x_0))$ on which the divisors on the selected path are nonzero and its strict comparisons retain their signs. Let $z_j$ be the test values that vanished on the path, taking the difference for a comparison of two scalars. Since squares of real numbers cannot cancel, a single scalar condition captures all these equalities. Set
\[
P=\sum_j z_j^2 .
\]
On $\mathcal N$, the condition $P=0$ says precisely that every $z_j$ vanishes. The input then passes each test on the path in order and reaches the same accepting leaf. Correctness of the tree implies $y=g(x)$. Conversely, all graph points over the open box follow this path, so $P(x,g(x))=0$ there. First choose a closed output box $Y$ with $g(x_0)$ in its interior. Then choose a sufficiently small closed input box $X$ inside the open box, with $x_0$ in its interior, so that $X\times Y\subset\mathcal N$ and the distance $\rho$ from $g(X)$ to $\partial Y$ is positive. For every fixed $x\in X$, the minimum of $P(x,\cdot)$ on all of $Y$ is zero and is attained only at $g(x)$.

\subsection{Perturbing the minimizer}

As in the scalar example where we subtracted $wc$, fix the same small vector $w\in\R^m$ for all $x$ and minimize $P(x,y)-w\cdot y$ over all of $Y$. Among the global minimizers choose one with the least first coordinate, then the least second coordinate among those remaining, and so on. Denote the resulting point by $h_w(x)$. This choice exists because the set of minimizers is nonempty and compact. Write $Q_{x,w}(y)=P(x,y)-w\cdot y$. A point $y$ is selected precisely when no competitor has a smaller objective value and no competitor with the same value precedes $y$ in lexicographic order. A single formula expresses both requirements.
\[
y\in Y\quad\land\quad
\neg\exists z\in Y\ \bigl[
 Q_{x,w}(z)<Q_{x,w}(y)\ \lor\
 (Q_{x,w}(z)=Q_{x,w}(y)\ \land\ z\prec_{\rm lex}y)\bigr].
\]
Here $z\prec_{\rm lex}y$ means that $z$ has the smaller coordinate at the first position where the two points differ. The box constraints and lexicographic order involve only finitely many polynomial sign conditions. Comparisons of rational functions defined on the box can also be written as polynomial conditions by clearing denominators with their squares. Applying real quantifier elimination to the whole formula eliminates the competitor $z$ while retaining the free variables $w,x,y$. The result describes the graph of the selection by a Boolean combination of signs from a single finite list of polynomials. Since we obtain this list before substituting a particular $w$, its length and degrees are bounded uniformly over all perturbations. In the next subsection the same argument will describe the points with large derivatives. A list valid for all perturbations will give a uniform bound on the number of bad intervals on a line.

Comparing the objective at $h_w(x)$ with its value at $g(x)$ gives
\[
0\le P(x,h_w(x))\le w\cdot\bigl(h_w(x)-g(x)\bigr)\le \diam(Y)\,\|w\| .
\]
For $0<\eta<\rho$, the continuous function $P$ has a strictly positive lower bound $\mu(\eta)$ on the compact set $\{(x,y)\in X\times Y:\|y-g(x)\|\ge\eta\}$. Hence $\diam(Y)\|w\|<\mu(\eta)$ implies $\sup_X\|h_w-g\|<\eta$. Thus $h_w$ converges uniformly to $g$. For small perturbations all selected minimizers lie in the interior of $Y$ and satisfy
\[
F\bigl(x,h_w(x)\bigr)=w,\qquad F:=\partial_yP .
\]
In the scalar example the equation had nonzero derivative at $c_w$. For several outputs the corresponding requirement is that the matrix $J=\partial_yF$ be invertible. To treat inputs and perturbations together, consider $\Phi(x,y)=(x,F(x,y))$ on an open neighborhood of $X\times Y$. Its Jacobian determinant is $\det\partial_yF$. Mark $(x,w)$ if some point $(x,y)$ in this neighborhood satisfies $F(x,y)=w$ and has singular $J$. The marked pairs are exactly the critical values of $\Phi$, a set of measure zero in $(x,w)$ by Sard's theorem. Now slice this set at fixed $w$. By Fubini's theorem the $x$-measures of the slices have integral zero over $w$, so almost every small $w$ has a slice of $x$-measure zero. Fix such a $w$ and take $x$ outside its exceptional slice. Then \emph{every} solution of $F(x,\cdot)=w$ in the neighborhood has invertible $J$, since a single singular solution would have marked $(x,w)$. Choose a sequence of such nonzero perturbations $w_k\to0$. At each regular solution the implicit function theorem gives a smooth local branch. The global minimizer is selected by its objective value and can switch branches as $x$ varies. To differentiate $h_w$ we must exclude these switches near the point we select. For fixed $w$, take a semialgebraic $C^2$ decomposition for $h_w$. The function is twice continuously differentiable on each full-dimensional open piece. The lower-dimensional pieces and the boundaries together lie in a closed null set \cite{Hardt}.

\subsection{From uniform approximation to approximation of mixed derivatives}

The values of $h_w$ are now close to the correct output. But the term $UV$ in the matrix expansion measures the joint effect of the two input changes and depends on a second derivative across the two input groups. A function may be uniformly small while its second derivative is large at a particular point. For example, $k^{-2}\cos(kz)$ converges uniformly to zero, yet its second derivative at $z=0$ is always $-1$. The next lemma estimates the volume on which a derivative of the error is large. On a line, a large derivative of constant sign cannot persist over a long interval. The common semialgebraic description of the family also bounds the number of such intervals. Together these facts leave room for one point at which all coefficients can be read.

\begin{lemma}\label{lem:2nd}
Let $X$ be a closed box and let $f:\Lambda\times X\to\R^m$ be jointly semialgebraic. Write $\delta(\lambda)=\sup_X\max_k|f_{\lambda,k}|$ and assume $\delta(\lambda)\le1$ for every $\lambda$. There is a single constant $C$, independent of $\lambda$, with the following property. For each $\lambda$, in the region where $f_\lambda$ is locally $C^2$, the set on which some first or second partial derivative of some component, including a mixed partial derivative, has absolute value greater than $2\sqrt{\delta(\lambda)}$ has measure at most $C\delta(\lambda)^{1/4}$.
\end{lemma}

\begin{proof}
Write $\delta=\delta(\lambda)$. The case $\delta=0$ is immediate, so assume $\delta>0$. Take the finite set of directions $\{e_i\}\cup\{(e_i+e_j)/\sqrt2:i<j\}$. Fix one direction $e$ and one component. Within the locally $C^2$ region, consider separately the sets on which the first or second derivative along $e$ is greater than $\tau>0$ or less than $-\tau$. Call these the bad sets. Derivatives and local regularity can be described by quantified conditions on difference quotients. Apply quantifier elimination once to the whole family, retaining $\lambda,\tau,x$ as variables, to obtain a fixed list of polynomials. Only now fix $\lambda,\tau$ and restrict $x$ to a line $z+te$. Each polynomial becomes univariate in $t$, of degree at most its original degree. Identically zero polynomials have constant sign. The total number of real roots of the remaining polynomials is at most the sum of the degrees in the list. All signs are constant between consecutive roots, and the roots themselves contribute no length. Thus to bound the length of a bad set on a line we need to count at most $K$ intervals, where $K$ is independent of the line, the parameter, and the threshold.

If the first derivative has constant sign and absolute value greater than $\tau$ on $[a,b]$, then $\tau(b-a)\le2\delta$, so the length is at most $2\delta/\tau$. For the second derivative choose the sign of $\phi=\pm f$ so that $\phi''>\tau$ along the interval. Then $\phi(t)-\tau t^2/2$ is convex, and the midpoint inequality gives
\[
\tau\,\frac{(b-a)^2}{8}\le\frac{\phi(a)+\phi(b)}2-\phi\Bigl(\frac{a+b}2\Bigr)\le2\delta,
\qquad\text{or equivalently}\qquad \frac{\tau(b-a)^2}4\le4\delta ,
\]
so each interval has length at most $4\sqrt{\delta/\tau}$. Integrate over parallel lines and take the union over the components and the finite set of directions. The bad set has measure at most $C'(\delta/\tau+\sqrt{\delta/\tau})$. Taking $\tau=\sqrt\delta$ bounds this by $C\delta^{1/4}$. Finally, the polarization identity
\[
\partial_i\partial_j=\partial^2_{(e_i+e_j)/\sqrt2}-\tfrac12(\partial_i^2+\partial_j^2),
\]
shows that every mixed derivative has absolute value at most $2\tau$ outside the bad set.
\end{proof}

Apply the lemma to $f_w=h_w-g$. Along the perturbations already chosen, $\delta_k=\delta(w_k)\to0$. For each sufficiently small selected perturbation $w_k$, the bad set has measure strictly less than the volume of $X$. Remove also the critical-value slice for this $w_k$, the set where $h_{w_k}$ is not locally $C^2$, and the boundary of the box. These additional sets have measure zero, so a point $x_k\in\operatorname{int}X$ remains. At this one point all output components and all required derivatives for this perturbation are controlled simultaneously. The next perturbation may use a different point. Since $\partial_{u_i}\partial_{v_j}g=T(e_i,e_j)$ everywhere, at the selected points we have, component by component,
\[
\bigl|\partial_{u_i}\partial_{v_j}h_{w_k}(x_k)-T(e_i,e_j)\bigr|\le2\sqrt{\delta_k}\to0 ,
\]
Thus the mixed derivative tensors converge to $T$.

\subsection{An $O(s)$ rank bound from the shared circuit}

At each selected point, differentiate once in a direction $U$ in the first input group and once in a direction $V$ in the second. Taken together, the resulting output components form a bilinear map in $U,V$. The preceding subsection shows that these maps converge to $T$. We must still decompose each one using only $O(s)$ product terms. The selected path supplies a circuit for $P$. The implicit equation expresses the mixed derivative of $h_{w_k}$ through derivatives of $F=\partial_yP$, which we can count on that circuit.

Fix $k$ and write $h=h_{w_k}$, $L=Dh(x_k)$, $a=(U,0)$, and $b=(0,V)$. At $x_k$ the matrix $J$ is invertible, $h$ is smooth nearby, and the identity $F(x,h(x))=w_k$ holds throughout a neighborhood of $x_k$. Differentiating twice gives
\[
D^2h[a,b]=-J^{-1}\,D^2F\bigl[(a,La),(b,Lb)\bigr].
\]
At this fixed point $L$ acts on the input linear forms in a rank decomposition, and $J^{-1}$ acts on its output vectors. Both linear transformations preserve the rank upper bound. At this point the linear maps can be absorbed into the forms and vectors of the decomposition. This is an existence argument for a rank bound. The cost $s$ counts the steps executed by the given verifier.

Suppose the path has $q$ arithmetic steps and $r$ tests, with $q+r\le s$. Reuse its intermediate values and split each division into a reciprocal and a multiplication. Each test contributes at most one subtraction, one squaring, and one addition to the sum. Thus the number $H$ of gates needed to construct $P$ satisfies $H\le 2q+3r\le3s$. One reverse differentiation pass, as in Baur--Strassen \cite{BS}, adds at most four gates per original gate. The forward and reverse passes together use at most $5H=O(s)$ gates and compute all of $F=\partial_yP$ at once.

To extract the cross terms of this rational circuit, substitute $(x_k,h(x_k))+\alpha(a,La)+\beta(b,Lb)$ at the input. Work in the ring with $\alpha^2=\beta^2=0$ and track the coefficient of $\alpha\beta$ gate by gate. Each gate value has the form $c_0+c_1\alpha+c_2\beta+c_3\alpha\beta$, where $c_1$ is linear in $U$ and $c_2$ is linear in $V$. The $\alpha\beta$ coefficient of a product is $c_0d_3+d_0c_3+c_1d_2+c_2d_1$. Thus a multiplication introduces at most two new products of a linear form in $U$ with a linear form in $V$. The $\alpha\beta$ coefficient of a reciprocal is $-d_3/d_0^2+2d_1d_2/d_0^3$, which introduces at most one. Addition, subtraction, and constants introduce none. The $\alpha\beta$ coefficients of the outputs are exactly the components of $D^2F[(a,La),(b,Lb)]$, and all outputs share the same list of products. Consequently $D^2h(x_k)[a,b]$ has real rank $O(s)$. Multiplying the three gate-count factors above gives the explicit constant in the theorem. Taking the limit established in the preceding subsection proves Theorem~\ref{thm:main}. \qed

\section{Fixed-size border rank and the verification exponent}

For any fixed $n$, the main theorem relates the verification cost to the border rank of the matrix multiplication tensor $M_n$. To obtain an arithmetic exponent as the size grows, we need to pass from approximate to exact decompositions. Bini's argument fixes the size and the approximation first, then interpolates tensor powers so that the overhead disappears when we take the exponent.

\begin{lemma}[Bini \cite{Bini}]\label{lem:bini}
If $\brank_\C M_n\le r$, then $n^{\omega_\C}\le r$.
\end{lemma}

\begin{proof}
Fix $n$. The map taking $r$ triples of factors to the sum of their rank-one tensors is polynomial. If the border rank is at most $r$, a sequence of such sums converges coefficientwise to $M_n$. Every polynomial vanishing on the image still vanishes at the limit, so $M_n$ lies in the complex Zariski closure of the image. Curve selection and a Laurent-series lifting give an arc in the same parameter space along which the sum is $M_n+O(t)$ \cite[Lemmas~15 and~16]{GSS}. If $M_n$ already lies in the image, constant parameters suffice. In general the individual factors may have poles. The negative powers cancel in the tensor sum, whose constant term is $M_n$. Choose an integer $M\ge0$ such that no factor coordinate contains a term of degree below $-M$. In each factor retain the terms through $t^{2M}$. Any discarded term, multiplied by the other two factors, has degree at least $1$, so the constant term of the tensor sum is unchanged. Multiplying each factor by $t^M$ gives a polynomial degeneration with the same $r$ summands. At this stage $h=3M$ and the degree $D$ of the left-hand side below are fixed.
\[
\sum_{l=1}^r a_l(t)\otimes b_l(t)\otimes c_l(t)=t^hM_n+O(t^{h+1}).
\]
The coordinate degree bound $D$ controls the number of interpolation points. At any one such point the rank is at most $r$, since evaluating the parameter still gives a sum of the same $r$ terms. The $k$th tensor power is a polynomial in $t$ of degree at most $kD$, and its coefficient of $t^{kh}$ is $M_n^{\otimes k}=M_{n^k}$. Interpolate at $kD+1$ points. Each evaluation has rank at most $r^k$, so $\rank M_{n^k}\le(kD+1)r^k$. Conversely, if $\rank M_m=R$, recursively applying this fixed-size exact decomposition to matrix blocks gives $\rank M_{m^j}\le R^j$. Other sizes can be padded with zeros up to the next power of $m$. It follows that $m^{\omega_\C}\le\rank M_m$ for every $m\ge2$. Equality is immediate for $m=1$. Therefore $n^{k\omega_\C}\le(kD+1)r^k$. With $n,D$ fixed, take $k$th roots and let $k\to\infty$ to obtain $n^{\omega_\C}\le r$.
\end{proof}

\begin{proof}[Proof of Corollary~\ref{cor:mm}]
Take $T=M_n$ and $p=q=m=n^2$. Theorem~\ref{thm:main} gives $\brank_\C M_n=O(s)$, and Lemma~\ref{lem:bini} then gives $s=\Omega(n^{\omega_\C})$. Both the geometric choices of points and the degeneration may depend on the size.

For the upper bound, given $\varepsilon>0$, choose $m$ with $\rank_\C M_m\le m^{\omega_\C+\varepsilon}$ and apply this complex bilinear algorithm recursively. Representing complex numbers by pairs of reals increases the cost by only a constant factor. This computes $AB$ in $O_\varepsilon(n^{\omega_\C+\varepsilon})$ steps, after which $n^2$ tests compare it with $C$.
\end{proof}

\section{Randomized verification and white-box identity testing}

\paragraph{The randomized and deterministic costs.} Measured in input scalars, the three matrices have total length $N=3n^2$. Run the Freivalds test from the introduction \cite{Freivalds} twice independently, rejecting if either trial fails. The total cost is $O(N)$ steps. A correct product always passes, and an incorrect product is accepted with probability at most $1/4$.

Corollary~\ref{cor:mm} identifies the verification exponent with $\omega_\C$ without assuming a particular value for it. Suppose in addition that we use the fixed-size matrix multiplication border-rank lower bound $n^{9/4}$ stated in the independent preprint of November 24, 2026 \cite{Pre}. Theorem~\ref{thm:main} then gives a deterministic lower bound of $\Omega(n^{9/4})=\Omega(N^{9/8})$. The ratio to the randomized cost above is at least $\Omega(N^{1/8})$. This numerical separation assumes the lower bound in that preprint. Both algorithms use polynomial time in this model.

\paragraph{White-box polynomial identity testing.} Matrix verification can also be expressed as a polynomial identity test. In both constructions below the decision tree receives the entries of $A,B,C$ as explicit real coefficients and is given the circuit using them. The first construction makes each entry of the residual the coefficient of a quadratic monomial. The second encodes each row of the residual as a univariate polynomial.

\begin{proposition}\label{prop:pit}
(i) Consider the fixed quadratic circuit $(x^\T A)(By)-x^\T Cy=x^\T(AB-C)y$. The coefficient of $x_iy_j$ is the corresponding entry of the residual. With $S=\Theta(n^2)$ coefficient slots, deterministic identity testing requires $\Omega(S^{\omega_\C/2})$ steps, whereas randomized evaluation takes $O(S)$ steps.

(ii) In the univariate case, the input is again the given circuit and its explicit coefficients. Set $v(t)=(1,t,\dots,t^{n-1})^\T$ and let
\[
Q(t)=\sum_i\bigl((A(Bv)-Cv)_i\bigr)^2 .
\]
The circuit computes $Bv$ once and reuses it for every row of $A$. Its size is $\Theta(n^2)$ and the degree is at most $2n-2$. This gives the same deterministic lower bound $\Omega(S^{\omega_\C/2})$.
\end{proposition}

\begin{proof}
In both cases the circuit computes the zero polynomial exactly when $AB=C$, so Corollary~\ref{cor:mm} gives the lower bounds. For the randomized test in (i), take independent uniform $0/1$ vectors $x,y$. If the residual $R\neq0$, then $Ry\neq0$ with probability at least $1/2$. Conditional on that event, $x^\T Ry\neq0$ with probability at least $1/2$. A constant number of independent repetitions brings the false acceptance probability below $1/4$, at a cost of $O(S)$ per trial. In (ii), $(A(Bv)-Cv)_i=\sum_j(AB-C)_{ij}t^{j-1}$. A sum of squares of real polynomials vanishes identically only if every residual row polynomial vanishes identically.
\end{proof}

The lower bound in Proposition~\ref{prop:pit}(ii) is for identity testing on the given circuit. If the input is the list of coefficients of $Q$ in expanded univariate form, it suffices to inspect those coefficients.

\paragraph{A matching upper bound for explicit dense quadratic inputs.} Suppose the input gives all coefficients of $K$ pairs of dense affine linear forms. We ask whether the sum of their products vanishes identically. The input length $S$ counts every coefficient slot, including those containing zero. After homogenization, each factor has a $d$-dimensional coefficient vector. Form a matrix $R\in\R^{d\times2K}$ whose first block of columns contains the vectors for the first factors and whose second block contains those for the second factors. Let $J=\begin{pmatrix}0&I_K\\I_K&0\end{pmatrix}$ exchange these blocks. Since $\sum_k(r_k\cdot z)(r'_k\cdot z)=z^\T(\sum_kr_kr_k'^\T)z$, the polynomial vanishes identically if and only if $RJR^\T=0$.

Write $c=2K$. If $d\le c$, partitioning along the inner dimension writes $(RJ)R^\T$ as a sum of $\lceil c/d\rceil$ matrix products of order $d$, padding the last block with zeros if necessary. If $d>c$, first compute $H=R^\T R$ in row blocks. Over the reals $\ker H=\ker R$, so taking orthogonal complements gives $\operatorname{im}H=\operatorname{im}R^\T$. The condition $RJR^\T=0$ says that the bilinear form defined by $J$ vanishes identically on this subspace. It is therefore enough to test $HJH=0$.

For $d\ge c$ the block computation costs $(d/c)\,c^{\omega_\C+o(1)}\le(dc)^{\omega_\C/2+o(1)}$. For $d\le c$ the same estimate holds with $d,c$ interchanged. Since $S=\Theta(dc)$, both cases give an upper bound of $S^{\omega_\C/2+o(1)}$. The circuit in Proposition~\ref{prop:pit}(i) has $n$ products from $(x^\T A)(By)$ and $n$ from $x^\T Cy$, all in the same $2n$ variables. Filling its unused coefficient slots with zeros therefore takes only $\Theta(n^2)$ slots. Thus the lower bound matches the upper bound in the exponent in the worst case for this input class.

\section{Prior work and scope}

Example~1 of the 1992 report by B\"urgisser, Karpinski, and Lickteig posed the problem of verifying the graph of real matrix multiplication. Their Example~2 gave a randomized $O(m)$ versus deterministic $\Omega(m\log m)$ separation for the graph of all elementary symmetric functions \cite{BKL}. Sections~3 and~4.2 of K\"unnemann's 2018 work \cite{K} already give bilinear encodings and a univariate reduction. Here we obtain real white-box time lower bounds that charge for evaluation. Bennett et al.\ \cite{Ben} gave algorithms for sparse residuals in 2024, as well as lower bounds for a restricted class of algorithms using independent projections. When arbitrarily large integers are allowed, Koiran's witness sequences give a polynomial arithmetic-time upper bound for identity testing of general division-free circuits \cite{Koi}.

The base field matters. Complex decision trees use equality tests. Simulating complex operations and equality tests with pairs of real numbers costs only a constant factor per step, with a corresponding change in the constant in the linear rank bound. Matrix multiplication has integer structure constants, so first-order transfer applies at each fixed size. The ordered model over the reals transfers to other real closed fields, and the equality-test model over the complex numbers transfers to other algebraically closed fields of characteristic zero.

For positive characteristic the order of the quantifiers is essential. First fix the matrix size $n$ and the cost $s$. If no corresponding equality-test tree of cost at most $s$ exists over the complex numbers, then none exists over an algebraically closed field of any sufficiently large prime characteristic. The threshold may depend on both $n$ and $s$. A general verification lower bound in fixed positive characteristic remains open. The univariate reduction in Proposition~\ref{prop:pit}(ii) also uses the sum-of-squares property over the reals.

\end{document}